\documentstyle[twocolumn]{jpsj}

\def\runtitle{Magnetic phase diagram and metal-insulator transition of NiS$_{2-x}$Se$_{x}$}

\def\runauthor{Masato {\sc Matsuura}, Haruhiro {\sc Hiraka}, Kazuyoshi {\sc Yamada} and Yasuo {\sc Endoh}}

\newcommand{\sub}[1]{\mbox{\scriptsize #1}}

\title
{
Magnetic Phase Diagram and Metal-Insulator Transition of NiS$_{2-x}$Se$_{x}$
}

\author
{ 
Masato {\sc Matsuura}\footnote{E-mail: matsuura@iiyo.phys.tohoku.ac.jp}, Haruhiro {\sc Hiraka},
Kazuyoshi {\sc Yamada}\footnote{Present address: 
Institute for Chemical Research, Kyoto University, Gokasho, Uji, Kyoto 611-0011.} and
Yasuo {\sc Endoh}
}

\inst
{
Department of Physics, Tohoku University, Sendai 980-8578
}

\recdate
{
December 22,1999
}

\abst
{
 Magnetic phase diagram of NiS$_{2-x}$Se$_{x}$ has been reexamined by systematic studies
 of electrical resistivity, uniform magnetic susceptibility and neutron diffraction
 using single crystals grown by a chemical transport method.
 The electrical resistivity and the uniform magnetic susceptibility exhibit 
 the same feature of temperature dependence over a wide Se concentration.
 A distinct first order metal-insulator (M-I) transition accompanied by a volume change
 was observed only in the antiferromagnetic ordered phase for $0.50\leq x\leq 0.59$. 
 In this region, the M-I transition makes substantial effects to the thermal evolution of staggered moments.
 In the paramagnetic phase,  the M-I transition becomes broad; both the electrical resistivity 
 and the uniform magnetic susceptibility exhibit 
 a broad maximum around the temperatures 
 on the M-I transition-line extrapolated to the paramagnetic phase.
 }

\kword
{
metal-insulator transition, mott-insulator, antiferromagnetism, neutron diffraction, phase diagram
}

\input epsf.sty

\begin{document}
\sloppy
\maketitle

\section{Introduction}
  
 NiS$_{2}$ has been recognized for a long time as a prototype of Mott insulator
 described by the Mott-Hubbard narrow band picture.
 Hence, many experimental results on magnetism and transport properties in this system
 have been interpreted by this scenario,~\cite{rf:wilson85}
 where a metallic phase appears by a broadening of 3d-bands 
 with either applying pressure beyond $\sim$3.5 GPa 
 or $\sim$25\% substitution of S sites with Se atoms.

 However, a recent photo-emission spectroscopy experiment revealed that 
 subsequent change in the band structure upon Se-substitution should not be described 
 by the band broadening but is governed by the charge-transfer within a band gap.~\cite{rf:fujimori96}
 Moreover, recent observations by the angle-resolved photoemission spectroscopy (ARPES)
 in the metallic phase near the phase boundary of metal-insulator transition (M-I boundary)
 showed a sharp peak in the spectrum near the Fermi energy.~\cite{rf:matsuura96,rf:matsuura98} 
 Enhancement of the effective mass near the M-I boundary ~\cite{rf:miyasaka99}
 also suggests a breakdown of the previous band picture.
 
 An antiferromagnetic (AF) long-range order has been confirmed to exist in both metallic and 
 insulating phases.~\cite{rf:gautier75}
 The AF long-range ordered structure in NiS$_{2}$ is complicated 
 showing two types of structure (${\bf q}_{\sub{M1}}=(001)$ and ${\bf q}_{\sub{M2}}=(\frac{1}{2}\frac{1}{2}\frac{1}{2})$ on fcc 
 lattice).~\cite{rf:hasting70,rf:miyadai75,rf:kikuchi78_1,rf:kikuchi78_2}
 Although several studies on magnetic phase diagram of NiS$_{2-x}$Se$_{x}$ were 
 carried out,~\cite{rf:jarrett73,rf:gautier75,rf:czjzek76,rf:sudo92,rf:yao97}
 there still exist non-trivial discrepancies in the N\'eel temperature ($T_{\sub{N}}$) 
 and M-I transition temperature ($T_{\sub{MI}}$).
 Furthermore, no systematic neutron study on NiS$_{2-x}$Se$_{x}$ with single crystal has been carried out.
 
 We succeeded in growing a series of single crystals over a wide Se-concentration.
 Then, we performed a systematic study to reexamine the magnetic phase diagram and the M-I transition itself,
 and established a new phase diagram of NiS$_{2-x}$Se$_{x}$ system with which
 an intimate relation between magnetic and transport properties was made it clear.
 
\section{Experimental Detail}
 Prescribed amount of Ni, S and Se powders in 5N purity were mixed with 3\%
 of excess S and Se. They were sintered in an evacuated silica tube at 720$^{\circ}\mbox{C}$
 for 7 days. 
 In order to improve the homogeneity of starting powders, 
 we repeated the above process of sintering and grinding twice.
 The Se-concentration of polycrystalline samples was determined from the lattice constant by utilizing 
 Vegard's law between NiS$_{2}$ (5.688\AA) and NiSSe (5.813\AA).~\cite{rf:gautier75}
 
 Then, single crystals were grown by a chemical transport method using Cl$_{2}$ gas.~\cite{rf:bouchard68}
 Polycrystalline powder of 0.5 g was sealed under 0.5 atm of Cl$_{2}$ gas
 in an evacuated silica ampoule (10$\phi$ in inner diameter and 14-20 cm in length)
 and was placed in a furnace. The average temperature was kept at about 750$^{\circ}\mbox{C}$
 with the temperature gradient, 2$^{\circ}\mbox{C}$/cm. 
 Single crystals were grown up to 4 mm on an edge for a month of growth. 
 They are characterized by shiny (100) and (111) facets.

 The electrical resistivity ($\rho$) was measured by the standard four-probe method 
 between 4.2 K and room temperature. 
 The uniform magnetic susceptibility ($\chi$) was measured with 
 a standard SQUID magnetometer under a magnetic field of 1 T from 5 K to 300 K. 
 Neutron diffraction measurements were performed on the triple-axis spectrometer TOPAN 
 installed at the JRR-3M Reactor of Japan Atomic Energy Research Institute. 
 Incident and final neutron energy was fixed to be 14.7 meV ($\lambda=2.67$ \AA$^{-1}$) 
 using pyrolytic graphite (PG) monochromator and analyzer. 
 Horizontal collimation of the neutron beam was set to be 60'-30'-60'-100',
 from the forefront of the monochromater to the entrance of neutron detector.
 The sample was mounted so as to access to the ({\it h,h,l}) reciprocal lattice plane.
 In this paper, we denote the indices of reflection in the AF unit cell 
 which is twice as large as the chemical unit cell,
 as used in previous works.~\cite{rf:sudo92}
 Temperature dependence of lattice constant was measured by using
 a standard four-circle spectrometer with x-rays provided from a rotating anode
 (40 kV$\times$100 meV, MoK$\alpha_{1}$) and monochromized using PG(002) reflection.
 
 In order to investigate the M-I transition in detail, we performed simultaneous 
 measurements on $\rho$ and staggered magnetization for $x=0.50$ and 0.53 by contacting copper wires 
 for resistivity measurement onto the crystal
 mounted in an aluminum can for the neutron diffraction measurement.
 Intensities of AF(002) Bragg 
 reflection (type I) and $\rho$ were monitored every 30 and 50 seconds, respectively.
 We changed the temperatures at a constant rate of  0.5 K/min.
 
\section{Experimental Results}
\subsection{Neutron diffraction}
 Bragg reflections corresponding to type I AF structure were observed in both the insulator 
 and the metallic phase ($x\geq 0.5$), while reflections of type II AF structure vanish for $x>0.3$,
 which is consistent with the previous result.~\cite{rf:miyadai83}
 Thermal evolutions of peak intensity of (002)  AF Bragg reflection $I$({\it T}),
 as typical examples of type I, are shown in Figs. 1 (a) for $x=0.50$ and (b) for $x=0.69$. 
\begin{figure}[hbtp]
 \begin{center}
  \begin{minipage}{86mm}
   \epsfxsize=86.0mm
   \epsfbox{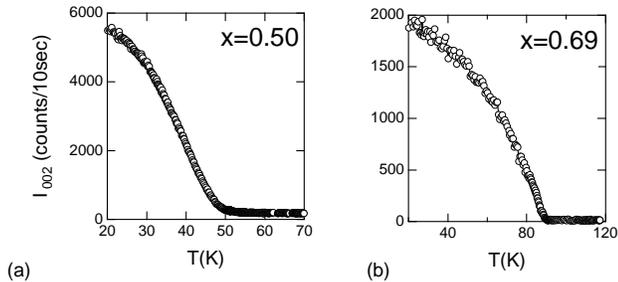}
   \caption{Thermal variation of peak intensity of (002) antiferromagnetic Bragg reflection 
   for NiS$_{2-x}$Se$_{x}$ (a) {\it x}=0.50 and (b) {\it x}=0.69}.
   \label{Figures 1}
  \end{minipage}
 \end{center}
\end{figure}

 We determined both $T_{\sub{N}}$ and a critical exponent $\beta$ on the basis of eqs.(1) and (2),
 assuming a Gaussian distribution of $T_{\sub{N}}$.
\begin{equation}
 I(T) \propto \left\{
 \begin{array}{ll}
 \int \left(\frac{|T-T_{\sub{N}}|}{T_{\sub{N}}}\right)^{2\beta}f(T_{\sub{N}}){\mib d}T_{\sub{N}} & (T<T_{\sub{N}})\\
 0 & (T>T_{\sub{N}}),
 \end{array}
 \right.
\end{equation}
\begin{equation}
 f(T_{\sub{N}})=\frac{1}{\sqrt{{2\pi}}\sigma}{\mbox{exp}}\left(-\frac{(T_{\sub{N0}}-T_{\sub{N}})^{2}}{2\sigma^{2}}\right).
\end{equation}
 The least-square fitting gives $T_{\sub{N0}}=45.0$ K, 
 $\sigma=7.4$ K and $\beta=0.71$ for $x=0.50$, 
 and $T_{\sub{N0}}=89.4$ K, $\sigma=0.9$ K and $\beta=0.39$ for $x=0.69$.
 The distribution of $T_{\sub{N}}$ $\sigma$($T_{\sub{N}}$)
 can be converted into the distribution of Se-concentration
 $\sigma$($\it x$) in the sample by using the $x$-dependence of $T_{\sub{N}}$.
 The $\sigma$($\it x$) is determined to be $\sim$ 0.01 for both $x=0.50$ and $x=0.69$.
 For samples with {\it x}$<$0.44 and {\it x}$>$0.60, 
 the critical exponent $\beta$ is determined to be 0.35$\sim$0.40,
 which is consistent with the theoretical value of the three dimensional Heisenberg model (0.367).
 On the other hand, for samples with 0.44$\leq${\it x}$\leq$0.59, $\beta$ exceeds 0.5.
 We note that the unusually large $\beta$ is observed only for samples 
 with $T_{\sub{N}}$ close to the M-I boundary.

 Assuming the noncollinear type I AF structure,~\cite{rf:kikuchi78_1,rf:kikuchi78_2} 
 we calculated the magnitude of sublattice magnetic moment  from the intensities of
 two AF (002), (220) and nuclear (222) reflections measured at 10K. 
 We used the diffraction data from polycrystalline powder samples 
 to minimize the uncertainty of extinction effect.
 The magnetic moment thus obtained is shown in Fig. 2.
 Our result almost reproduces previous one by Miyadai {\it et al}.~\cite{rf:miyadai83}
 However, our data indicate a change in Se-concentration dependence around the M-I boundary;
 the staggered moment decreases more rapidly with Se-substitution in the metallic phase 
 than in the insulating one.\\ 
\begin{figure}[hbtp]
 \begin{center}
  \begin{minipage}{60mm}
    \epsfxsize=60.0mm
    \epsfbox{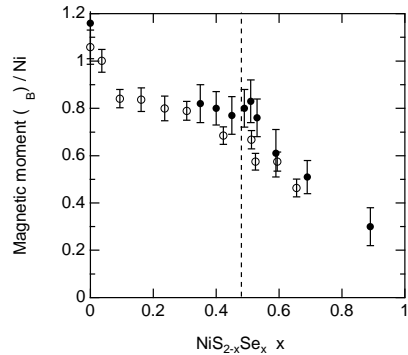}
    \caption{{\it x}-dependence of the magnetic moment for NiS$_{2-x}$Se$_{x}$ at 10 K. 
	Closed and open circles represent 
    present data and those by Miyadai {\it et al}.,~\cite{rf:miyadai83} respectively. 
	Dotted line shows the M-I boundary at 0 K.}
    \label{Figure 2}
  \end{minipage}
 \end{center}
\end{figure}

\subsection{Electrical resistivity and uniform magnetic susceptibility}
 The temperature dependences of $\rho$ and $\chi$ are shown in Figs. 3 and 4, respectively. 
 We note that both measurements were performed for the identical crystals. 
 From thermal evolutions of $\rho$ and $\chi$, we can categorize 
 four compositional regions in terms of M-I transition;
 (1) semiconducting region($0\leq x\leq0.47$), (2) first-order M-I transition ($0.50\leq x\leq 0.59$),
 (3) broad M-I transition ($x \sim 0.65$), and (4) metallic region($x\geq 0.69$).\\
 (1) Semiconducting region ($0\leq x\leq 0.47$)
 
 $\rho$ exhibits a typical feature of activation type for usual  semiconductors.
\begin{figure}[thbp]   
 \begin{center}
  \begin{minipage}{85mm}
   \begin{center}
    \epsfxsize=85mm
    \epsfbox{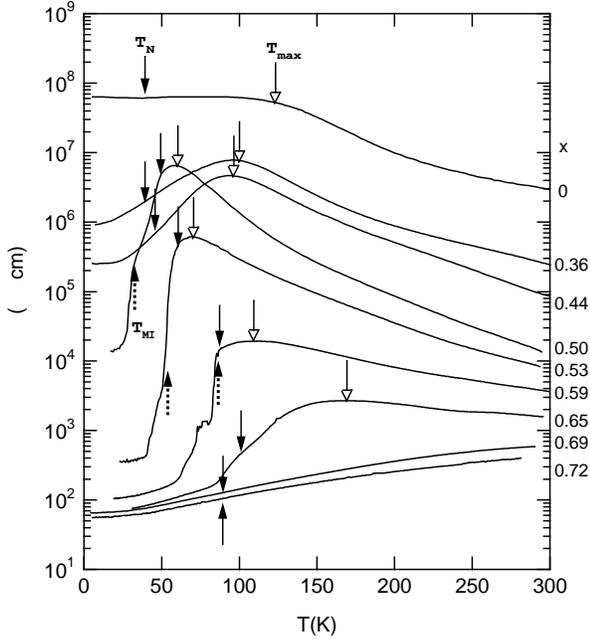}
    \caption{Temperature dependences of the electrical resistivity ($\rho$) 
	 for NiS$_{2-x}$Se$_{x}$ single crystals. 
     Solid arrows indicate $T_{\mbox{\tiny N}}$ determined by neutron diffraction measurements.
     Dashed arrows point at the temperature where $\rho$ drops discontinuously.
     Open arrows represent the temperature at which $\rho$ reaches the maximum.} 
    \label{Figure 3}
   \end{center}
  \end{minipage}
 \end{center}
\end{figure}
 The activation energy $E_{\sub{a}}$ at room temperature decreases with increasing Se-substitution
 from 80 meV (NiS$_{2}$) to 50 meV ($x=0.47$).
 These values are consistent with the hopping energies determined by Kwizera {\it et al}.~\cite{rf:kwizera80}

 $\chi$ shows a well-defined cusp corresponding to $T_{\sub{N}}$ 
 determined by neutron diffraction measurements. 
 $T_{\sub{N}}$ is almost constant at 40 K in the semiconducting region.
 For NiS$_{2}$, a weak ferromagnetism appears below $T_{\sub{c}}=30$ K.
 For samples with $0.36\leq x\leq 0.47$, 
 $T_{\sub{c}}$ drops down to 20 K with the weak bulk ferromagnetization, 
 which is 10$^{-3}$ times smaller than that of NiS$_{2}$.
 
 $\rho$ exhibits a plateau or a broad maximum at $T_{\sub{max}}$ 
 as indicated by open arrows in Fig. 3.
 $T_{\sub{max}}$ is higher than $T_{\sub{N}}$ and decreases with increasing {\it x} 
 from 120 K (NiS$_{2}$) to 80 K ($x=0.47$).
 $\chi$ does not follow a simple Curie-Weiss law above $T_{\sub{N}}$.
 These facts suggest that the short-range magnetic correlation develops well above $T_{\sub{N}}$,
 strongly coupled with transport properties.\\
 (2) First order M-I transition ($0.50\leq x\leq 0.59$)
 
 At high temperatures, $\rho$ shows an activation type behavior.
 $E_{a}$ determined at room temperature decreases with Se-substitution from 30 meV ($x=0.50$)
 to 20 meV ($x=0.59$).
\begin{figure}[thbp]
 \begin{center}
  \begin{minipage}{85mm}
    \epsfxsize=85mm
    \epsfbox{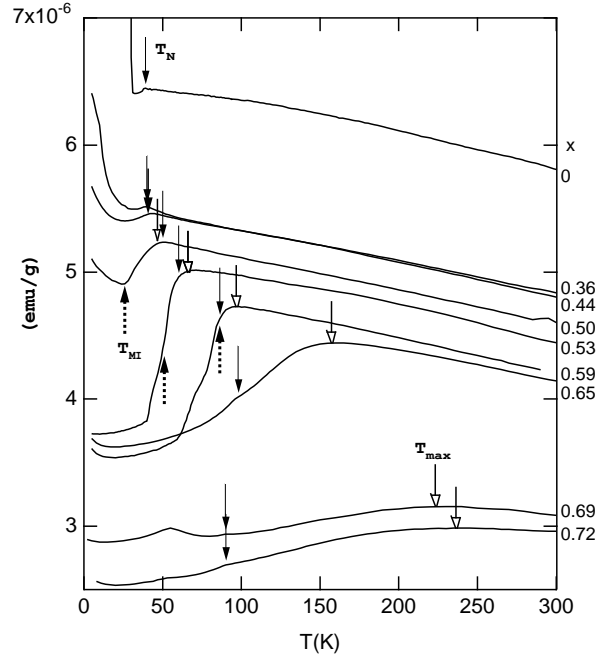}
    \caption{Temperature dependences of the uniform magnetic susceptibility ($\chi$) for NiS$_{2-x}$Se$_{x}$ 
	single crystals. 
    Solid and dashed arrows indicate $T_{\mbox{\tiny N}}$ and $T_{\mbox{\tiny MI}}$, as in Fig.3.
    Open arrows represent the temperature at which $\chi$ reaches the maximum.} 
    \label{Figure 4}
  \end{minipage}
 \end{center}
\end{figure}

 The whole process of the M-I transition in NiS$_{2-x}$Se$_{x}$ seems to occur in a broad temperatures
 range with a width of 20$\sim$30 K.
 Both $\rho$ and $\chi$ exhibit a maximum (open arrows in Figs. 3 and 4) 
 at temperatures higher than $T_{\sub{N}}$ (solid arrow), 
 but no cusp appears at $T_{\sub{N}}$ in $\chi$. 
 Below $T_{\sub{N}}$, $\rho$ rapidly decreases with decreasing temperature, 
 and then further drops discontinuously by order of $10^{-1}\sim 10^{-2}$,
 indicating the first order M-I transition occurs in the AF ordered state.
 $T_{\sub{MI}}$ is defined as the temperature where $\rho$ discontinuously drops. 
 The difference between $T_{\sub{MI}}$ and $T_{\sub{N}}$ becomes small with increasing {\it x}.

 $\rho$ further decreases below $T_{\sub{MI}}$,
 and eventually exhibits a weak temperature-dependent metallic behavior.
 We have investigated the detailed feature of the M-I transition 
 by simultaneous measurements on the electrical resistivity 
 and the staggered magnetization, as described in $\S${\it 3.3}.\\
 (3) Broad M-I transition ($x \sim 0.65$)
 
 Although the phase transition from semiconductor to metal occurs on cooling, 
 no discontinuous change was observed in $\rho$ and $\chi$. 
 As shown for $x=0.65$ in Figs. 3 and 4, 
 both $\rho$ and $\chi$ exhibit a broad maximum in the paramagnetic phase;
 at the temperatures higher than $T_{\sub{N}}$ by about 60 K.
 $T_{\sub{max}}$ increases with increasing {\it x}, while $T_{\sub{N}}$ saturates in this region.
 In contrast to the region (2), $\chi$ shows a small cusp at $T_{\sub{N}}$.
 $\rho$ exhibits metallic behavior at low temperatures.
 It is noted that this broad M-I transition crossovers to the following metallic one 
 at the Se-concentration with the maximum  $T_{\sub{N}}$.\\
 (4) Metallic region ($x\geq 0.69$)
 
 $\rho$ shows a typical metallic behavior below 300 K as shown in Fig. 3 for $x\geq 0.69$. 
 However, a broad maximum in $\chi$ still exists well above $T_{\sub{N}}$ as indicated by 
 open arrows in Fig. 4, which suggests the persistence of spin correlation at much higher 
 temperatures above $T_{\sub{N}}$.
 Similar to the  previous work,~\cite{rf:ogawa77} 
 $T^{2}$ dependence in $\rho$ was observed at low temperatures.
 Crossover from $T^{2}$  to T-linear behavior occurs around temperatures where $\chi$ shows a maximum.

\subsection{Detailed simultaneous scan on electrical resistivity and neutron diffraction near M-I transition}
 
 As described above, the first order M-I transition 
 was observed for 0.50$\leq${\it x}$\leq$0.59 in the AF ordered phase. 
 In order to study correlation between the M-I transition and the magnetism in more detail,
 we performed simultaneous measurements on the electrical resistivity and staggered magnetization;
 the peak height of (002) AF Bragg reflection $I_{\sub{002}}$.
 Temperature dependences of $\rho$ and $I_{\sub{002}}$ measured simultaneously 
 are shown in Figs. 5 for (a) $x=0.50$ and (b) $x=0.53$ single crystals. 

 For both $x=0.50$ and 0.53, $\rho$ sharply drops below $T_{\sub{N}}$ and
 the first order M-I transition occurs in the AF ordering state.
 For $x=0.59$ as shown in Fig. 5(c), both magnetic order and metallic phase appear almost simultaneously.
 As is clearly seen in Figs. 5, 
 there exists an unusual correlation between the staggered magnetization and electrical resistivity.
 $\rho$  gradually decreases associated with tailing in $I_{\sub{002}}$ around $T_{\sub{N}}$,
 which gives unusually large critical exponent $\beta$.
 In addition, $I_{\sub{002}}$ show a small but clear kink at $T_{\sub{MI}}$ 
 where $\rho$ drastically change in Figs. 5(b) and 5(c).
 Moreover, there exists an unusual kink in $I_{\sub{002}}$ at temperature
 below which $\rho$ and $\chi$ show metallic behavior. 
 Below the kink temperature, $T_{\sub{kink}}$, temperature dependence of $I_{\sub{002}}$ becomes weak
 in metallic phase.  
 This phenomenon is not due to the extinction for the single crystal, 
 since the same phenomenon was also observed for $x=0.53$ in the polycrystalline powder sample.
\begin{figure}[thbp]
 \begin{center}
  \begin{minipage}{80mm}
   \epsfxsize=80.0mm
   \epsfbox{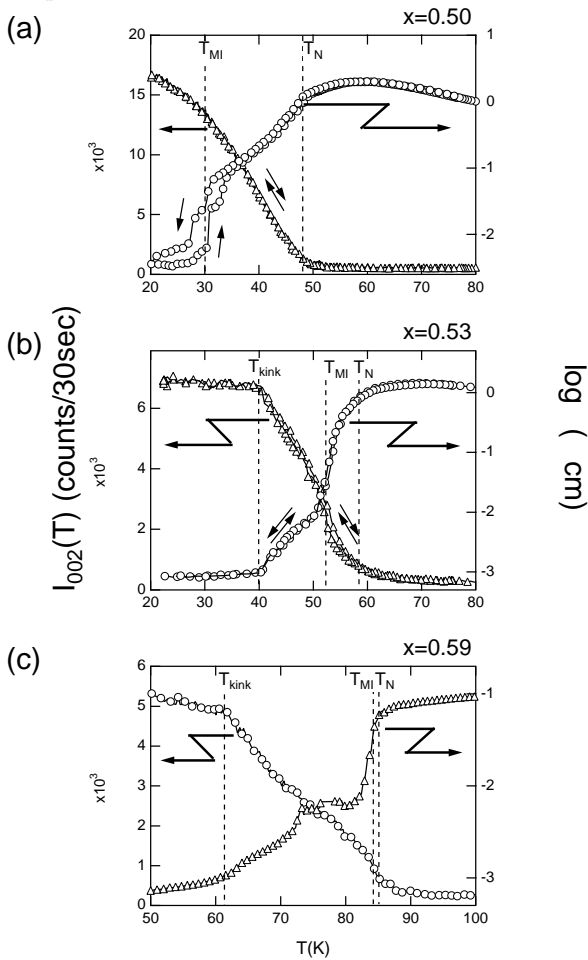}
   \caption{Temperature dependences of the electrical resistivity and
   the peak intensity of (002) antiferromagnetic Bragg reflection, $I_{\mbox{\tiny 002}}$({\it T})
   measured simultaneously for (a) $x=0.50$ and (b) $x=0.53$, and separately for (c) $x=0.59$.
   Circles and triangles indicate electrical resistivity and $I_{\mbox{\tiny 002}}$({\it T}), respectively.}
   \label{Figures 5}
  \end{minipage}
 \end{center}
\end{figure}

 It is noted $\rho$ shows a multi step change near the M-I transition as shown in Figs. 5.
 Here we describe on the multi-step like change in $\rho$.
 For the data from $x=0.50$, the multi step change near 30 K in Fig. 5(a) was concluded to be ascribed
 in the following extrinsic reason.
 The single crystals of $x=0.50$ were grown from a mixing of two powder samples 
 synthesized separately, which gives rise to a slightly difference in concentration.
 After this simultaneous measurement, we synthesized the polycrystalline powder sample of $x=0.50$ 
 and grew again the single crystal which shows a single step change at the M-I transition. 
 Therefore, we concluded the single crystal of $x=0.50$ showing two step anomalies 
 contained two regions with slightly different Se-concentration by about $x=0.005$.
 
 However, for x=0.53 and 0.59 we found no distinct experimental reason 
 to induce inhomogeneity of Se-concentration.
 The same multi step change of $\rho$ was also observed
 in a previous independent study on single crystals grown by a flux method.~\cite{rf:yao96} 
 Therefore, the multi-step change for x=0.53 and 0.59
 can be intrinsic unlike the present data for $x=0.50$.
 We only speculate that between $T_{\sub{kink}}$ and $T_{\sub{MI}}$,
 the metallic and the insulating phases possibly coexist 
 due to the first-order transition at $T_{\sub{MI}}$. 
 The multi step change of $\rho$ may relate with such an inhomogeneous state in this temperature region.
 
 As shown in Fig. 6, the lattice constant discontinuously changes 
 with a small thermal hysteresis only at the M-I transition, but no anomaly at $T_{\sub{kink}}$.
 The small difference in magnitude of lattice constant between Fig.6 and its inset is 
 due to the slight shift in position of sample, $0.02^{\circ}$, on cooling process.
 This first order M-I transition can be interpreted to be associated with 
 the volume contractions with decreasing temperature.
 Applying the Clausius-Clapeyron relation to the present data, entropy change at the M-I transition 
 $\Delta S$ is calculated to be 1.6 ($\pm 0.3$) [J/(mol deg.)], as previous report 
 of M-I transition induced by pressure.~\cite{rf:mori78}
 This value is in good agreement with that from the specific heat measurement by Sudo {\it et al}.,
 $\Delta S=1.0$ [J/(mol deg.)]  for $x=0.51$ polycrystalline powder sample. ~\cite{rf:sudo86}
 
\begin{figure}[hbtp]
\begin{center}
 \begin{minipage}{60mm}
  \epsfxsize=60.0mm
  \epsfbox{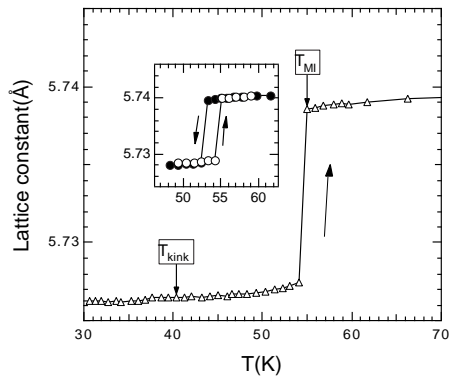}
   \caption{Temperature dependence of lattice constant for $x=0.53$.
   Open and closed circles represent data on heating and cooling processes, respectively.
   }
   \label{Figure 6}
  \end{minipage}
 \end{center}
\end{figure}

\section{Discussion}
 In Fig. 7, we present a revised phase diagram of NiS$_{2-x}$Se$_{x}$ as the summary of the present study.
 Here, we are mainly concerned about the phases near the M-I phase boundary for $x>0.3$.
 Compared with previous results of the phase diagram,~\cite{rf:jarrett73,rf:gautier75,rf:czjzek76,rf:sudo92,rf:yao97}
 we add two crossover regions as shown in Fig. 7.
 In the lower Se-concentration range, the crossover region depicted by vertical lines in the figure
 is determined by the temperature variation of transport properties. 
 For example, $\rho$ exhibits a broad maximum at the crossover region.
 Such a crossover was also observed in thermoelectric power measurements.~\cite{rf:kwizera80,rf:yao96}
 Furthermore, our recent neutron inelastic scattering data for NiS$_{2}$
 show that the short-range magnetic correlation remains above $T_{\sub{N}}$ at least up to $\sim$ 200 K.
 Therefore, we conjecture that the crossover behavior 
 in the transport properties is inherently correlated with the magnetic short-range correlation.
 We call the region below the crossover as 
 `short-range antiferromagnetic insulating (AFI)' phase.
 The onset of short-range AFI phase decreases with increasing {\it x}, 
 while $T_{\sub{N}}$ is almost constant in the insulating phase.

 Another crossover in the higher Se-concentration range approximately corresponds to the M-I transition-line 
 extrapolated towards the paramagnetic phase. 
 Around the crossover temperature $\rho$ and $\chi$ show a broad peak.
 For $x\geq0.69$, the peak in $\rho$ seems to disappear.
 For $0.50\leq x\leq 0.59$, the peak appears at temperature 
 slightly higher than that of the first order M-I transition.
 Here, we add a new phase called anomalous metal between $T_{\sub{N}}$ and $T_{\sub{max}}$.
 In Fig. 7, we refer the data by Ogawa for $x\geq 1.0$.~\cite{rf:ogawa79} 
  We note that such a broad maximum in $\chi$ was also observed 
 in the paramagnetic phase of La$_{2-x}$Sr$_{x}$CuO$_{4}$, 
 high temperature superconducting cuprates.~\cite{rf:nakano94}

\begin{figure}[thbp]
 \begin{center}
  \begin{minipage}{80mm}
   \epsfxsize=80.0mm
   \epsfbox{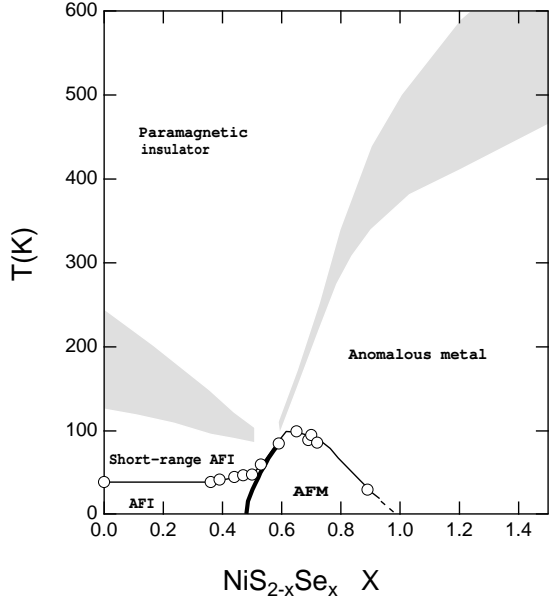}
   \caption{Magnetic phase diagram of NiS$_{2-x}$Se$_{x}$. 
   Open circles represent $T_{\mbox{\tiny N}}$. 
   Bold line shows a line of first order M-I transition (see Fig.8 for more detail).
   Lines are guide to the eyes.
   The region depicted by vertical lines indicate a crossover region 
   where transport properties or $\chi$({\it T}) reach a maximum; 
   $\rho$({\it T}) and Seebeck coefficient~\cite{rf:kwizera80,rf:yao96} for $x\leq 0.47$, 
   $\rho$({\it T}) and $\chi$({\it T}) for $0.50\leq x\leq 0.65$, and
   $\chi$({\it T}) for $x\geq 0.69$.
   For $x\geq 1.0$, we referred the data by Ogawa.~\cite{rf:ogawa79}
   AFI and AFM denote antiferromagnetic insulator and metal, respectively.
   }
  \label{Figure 7}
 \end{minipage}
\end{center}
\end{figure}
\begin{figure}[thbp]
\begin{center}
 \begin{minipage}{80mm}
  \epsfxsize=80.0mm
  \epsfbox{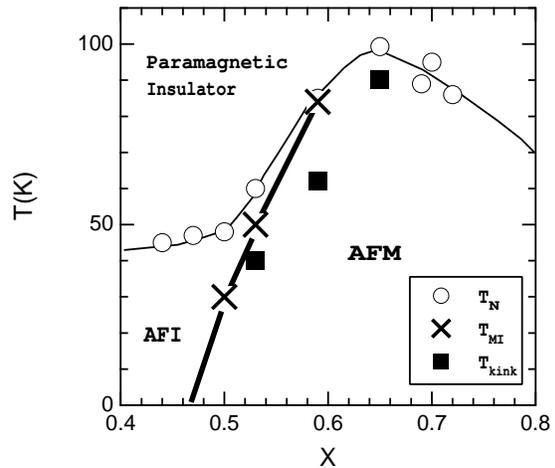}
  \caption{
  Detailed magnetic phase diagram  near M-I boundary.
  Solid squares point at temperatures where $\rho$({\it T}) shows a kink.
  Well-defined first order M-I transition was observed at $T_{\mbox{\tiny MI}}$, 
  represented by $\times$.
  }
  \label{Figure 8}
 \end{minipage}
\end{center}
\end{figure}

 A detailed magnetic phase diagram near the M-I boundary is shown in Fig. 8.
 We confirmed that the first order M-I transition occurs only in the AF ordered phase 
 for $0.50\leq x\leq 0.59$ from the simultaneous scan on $\rho$ and neutron diffraction,
 and the M-I transition becomes second order in the paramagnetic phase,
 as is seen in $\rho$ and $\chi$.
 According to the recent band calculation by the two-band Hubbard model in infinite dimension,
 a first order M-I transition in long range AF ordered phase is expected to occur
 in consideration of the electron hopping between nearest neighbor Ni 3d  and ligand 3p.~\cite{rf:watanabe98}
 It suggests the charge-transfer nature being indispensable 
 to understand the mechanism of M-I transition in this system. We note that the charge-transfer nature
 is seen in ARPES spectrum as a sharp peak near the Fermi energy in the metallic phase around the M-I boundary.
 
 It is important to compare the M-I transition in NiS$_{2-x}$Se$_{x}$ studied here with
 the previous experiment under pressure for NiS$_{2}$.
 Both phase diagrams under pressure and  Se-substitution are qualitatively 
 the same with each other.~\cite{rf:mori83}
 However, there exists an important difference between the two. 
 In fact, a recent electrical resistivity measurement revealed that
 the first order M-I transition in NiS$_{2}$ occurs at 150 K at 3.0 GPa,~\cite{rf:sekine97} 
 which can be converted into $x \sim 0.6$ for Se-substitution.
 On the other hand for Se-substitution, the maximum temperature for 
 both the first order M-I transition and the long range AF ordering are limited to 100 K. 
 Therefore, the observation of the M-I transition at 150 K under pressure may conflict with our conjecture
 that the first order M-I transition occurs only in the antiferromagnetic phase. 
 We speculate however, local randomness by Se-substitution may reduce the transition temperatures. 
 Therefore, it is highly required to determine the $T_{\sub{N}}$ under high pressure 
 beyond $\sim$3.5 GPa for NiS$_{2}$.
 
 The revised phase diagram in NiS$_{2-x}$Se$_{x}$ is quite similar to 
 those of other strongly correlated electron systems,
 such as V$_{2}$O$_{3}$,~\cite{rf:V2O3}
 and organic conductors (BEDT-TTF)$_{2}$X,~\cite{rf:BEDT-TTF1,rf:BEDT-TTF2,rf:BEDT-TTF3} 
 where antiferromagnetism and M-I transition are closely connected.
 In the latter material, a superconducting phase appears, 
 while V$_{2}$O$_{3}$ and NiS$_{2-x}$Se$_{x}$ become AF metals.
 
 In these systems, it is often observed that the onset temperature of long range ordered state 
 is significantly suppressed compared to the thermal robustness of short-range magnetic correlation.
 This is the case for NiS$_{2-x}$Se$_{x}$. Since Ni$^{2+}$ ions form an fcc lattice in this structure, 
 the nearest neighbor spins can be magnetically frustrated. In fact, preliminary inelastic neutron scattering 
 measurement on this system found an unusual magnetic excitations possibly due to this frustration effect.
 
\section{Summary}

 NiS$_{2-x}$Se$_{x}$ system has been reinvestigated systematically by neutron diffraction, 
 electrical resistivity, uniform magnetic susceptibility and X-ray diffraction 
 using both single crystals and powder samples. 
 We revealed a clear correlation between the transport and magnetic properties in this system. 
 Both the magnitude of magnetic moment and its temperature dependence are affected by electric 
 transport properties, particularly by the M-I transition.
 A well-defined first order M-I transition was found to occur only in the antiferromagnetic state. 
 On the other hand, the transition in the paramagnetic phase becomes broad 
 and is seen in the resistivity up to the Se-concentration where $T_{\sub{N}}$ exhibits a maximum.
 The novel feature in the newly proposed magnetic phase diagram is the existence of two cross-over regions, 
 one from short ranged AF phase to paramagnetic insulator (PI) and the other from anomalous metal to PI.
 These crossover features suggest that the short-range magnetic correlation remains well above $T_{\sub{N}}$
 in both phases, which cannot be interpreted by a single Mott Hubbard model,
 but by a new model taking account of the strong electron correlations.
 
\section*{Acknowledgements}
We would like to thank T. Takahashi and M. Onodera for their assistance in crystal growth, 
and H. Kimura, K. Hirota and K. Nemoto for their technical assistance in x-ray diffraction measurements and
neutron scattering.
This work was supported by Research Fellowships of the Japan Society for the Promotion of Science for 
Young Scientists.

\end{document}